\documentclass{PoS}

\usepackage{esvect}

\usepackage{gensymb}

\usepackage{multirow}

\usepackage{amsmath}

\title{TeV Observations of the Galactic Plane with HAWC and Joint Analysis of GeV Data from Fermi}

\ShortTitle{HAWC-Fermi Joint Analysis}

\author{\speaker{Hao Zhou}$^a$, C. Michelle Hui$^a$, and Petra Huentemeyer$^a$ for the HAWC Collaboration$^b$ \\
        \llap{$^a$}Department of Physics, Michigan Technological University, Houghton, MI, USA \\
        \llap{$^b$}For a complete author list, see \href{http://www.hawc-observatory.org/collaboration/icrc2015.php}{www.hawc-observatory.org/collaboration/icrc2015.php}. \\
        Email: \email{hzhou1@mtu.edu}}

\abstract{A number of Galactic sources emit GeV-TeV gamma rays that are produced through leptonic and/or hadronic mechanisms. Spectral analysis in this energy range is crucial in order to understand the emission mechanisms. The HAWC Gamma-Ray Observatory, with a large field of view and location at $19^\circ$ N latitude, is surveying the Galactic Plane from high Galactic longitudes down to near the Galactic Center. Data taken with partially-constructed HAWC array in 2013-2014 exhibit TeV gamma-ray emission along the Galactic Plane. A high-level analysis likelihood framework for HAWC, also presented at this meeting, has been developed concurrently with the Multi-Mission Maximum Likelihood (3ML) architecture to deconvolve the Galactic sources and to perform multi-instrument analysis. It has been tested on early HAWC data and the same method will be applied on HAWC data with the full array. I will present preliminary results on Galactic sources from TeV observations with HAWC and from joint analysis on Fermi and HAWC data in GeV-TeV energy range.}

\FullConference{The 34th International Cosmic Ray Conference,\\
		30 July- 6 August, 2015\\
		The Hague, The Netherlands}

\begin{document}

\section{Introduction}

GeV-TeV gamma rays are produced by energetic particles interacting with other particles or magnetic field. Supernova remnants (SNRs) and pulsar wind nebulae (PWNe) are most common Galactic sources in the TeV energy range. In PWNe, electrons are accelerated by the termination shock where the pulsar wind collides with the surrounding gas, emitting gamma rays via inverse Compton scattering. In SNRs, which are postulated as the cosmic-ray acceleration site, hadrons gain energy via diffuse shock acceleration, emitting gamma rays via pion decay. About half of TeV Galactic sources are still unidentified (UID) without associating in other wavelengths. Observations of GeV-TeV gamma rays are the key to distinguish between leptonic and hadronic processes which will aid in source identification.

The inner Galaxy region contains the strongest TeV Galactic sources other than the Crab Nebula. H.E.S.S. imaging atmospheric Cherenkov telescopes (IACTs) \cite{HESS:2006,HESS:2013} surveyed the Galactic plane in the southern sky and air shower arrays Milagro \cite{Milagro:2007} and ARGO \cite{Argo:2013} surveyed the Galactic plane in the northern sky. Targeted observations have been performed by IACTs H.E.S.S., VERITAS, and MAGIC (for example, \cite{MAGIC:2006,VERITAS:2009}). \textit{Fermi} Large Area Telescope (\textit{Fermi}-LAT) performed an unbiased all sky survey in the energy range of 20\,MeV-300\,GeV. Their third source catalog \cite{3fgl} is published based on the first four years of science data, and another catalog of sources above 10\,GeV is published with the first three years of \textit{Fermi}-LAT data \cite{1fhl}.

The High Altitude Water Cherenkov (HAWC) Gamma-Ray Observatory, located at Sierra Negra, Mexico at 4100\,m a.s.l., is sensitive to gamma rays in the energy range of 100\,GeV to 100\,TeV \cite{HAWC:2013}.  HAWC consists of an array of 300 water Cherenkov detectors (WCDs), covering an area of 22,000\,$\text{m}^2$.  Each WCD is 7.3\,m in diameter, 4.5\,m in depth, and filled with $\sim200,000$\,L of purified water.  Four upward-facing photomultiplier tubes (PMTs) are attached to the bottom of each WCD to detect the Cherenkov radiation produced by the secondary particles in an air shower. One high-quantum efficiency 10-inch Hamamatsu R7081-MOD PMT is in the center of each WCD and three 8-inch Hamamatsu R5912 PMTs are at 1.8\,m from the center \cite{Pretz:2015}.  The HAWC observatory has a high duty cycle of $>95$\% and a large instantaneous field of view of 2\,sr. 

A major challenge analyzing the emission from the inner Galaxy region, where has high potential of source confusion,  is to deconvolute and identify sources due to limited angular resolution of TeV air shower arrays. A likelihood framework \cite{Patrick:2015} has been developed to simultaneously fit the positions and fluxes of multiple sources, allowing to determine the number of sources in a region of interest (ROI). In these proceedings we present a TeV gamma-ray survey of the inner Galaxy region of $l\in [+15^\circ,\,+50^\circ]$ and $b\in[-4^\circ,\,+4^\circ]$ using the data taken with a partially-completed HAWC array, followed by a joint analysis using \textit{Fermi}-LAT data.

\section{Data and Analysis}

The full HAWC array was completed in March 2015, but science operations already started in August 2013 with a partially-built array. In these proceedings, the analysis is performed on data taken with the partially-constructed HAWC array, hereafter referred to as HAWC-111, between August 2, 2013 and July 8, 2014. During this construction phase, the active WCDs grew from 108 (362 PMTs) to 134 (491 PMTs). Data taking was occasionally interrupted during this period because the deployment of the detectors took precedence.
The dataset contains $275\pm1$ source transits in the inner Galaxy region.

The energy of a primary gamma-ray particle is correlated to the size of the air shower. Data are binned in 10 bins according to the fraction \textit{f} of PMTs that are triggered in an air shower with respect to the total live PMTs. HAWC-111 detects air shower events at a rate of 15\,kHz. However, $>99\%$ of air showers that trigger the detector are induced by charged cosmic rays, which are the main background for gamma-ray observations. A series of cuts are applied to the data \cite{Paco:2015} to remove cosmic ray events. Cuts are optimized in each \textit{f} bin to maximize the significance of the Crab nebula, which is the brightest steady source and a standard candle in the TeV energy range. 

The arrival direction of each air shower event is reconstructed using the relative time of the PMT signals when the air shower crosses the array \cite{Calibration:2015}. The arrival directions are then filled into a signal map using HEALPIX pixelation scheme \cite{Healpix:2005}. The average pixel scale is $0.11^\circ$, which is smaller than the angular resolution of the HAWC-111 detector. The background is estimated from the data using the method of direct integration \cite{Milagro:2003}. The integration time of two hours is chosen to emphasize the sources smaller than $30^\circ$ in angular size.

A set of Monte Carlo simulations, which simulate the detector response to gamma rays and cosmic rays, are used in the maximum likelihood method. These simulations are generated using the CORSIKA package \cite{Corsika:1998}, which simulates the shower propagation in the atmosphere and the GEANT4 package \cite{Geant4:2003}, which simulates the response of WCDs to air shower particles.

\subsection{The Likelihood Method}

The likelihood method begins with building a source model. A point source is characterized by its position and spectrum. 
The position is described by right ascension and declination. The spectrum is given by a simple power law 

\begin{equation}
\frac{dN}{dE} = I_0 \left(\frac{E}{E_0}\right)^{-\alpha}, 
\end{equation}

\noindent where $I_0$ is the flux normalization, $E_0$ is the pivot energy, and $\alpha$ is the spectral index.
The study presented here assumes a fixed index of 2.3 according to the average measured spectral index for known Galactic objects \cite{HESS:2006}. 
 
In each pixel of each \textit{f} bin, the observed number of events follows a Poisson distribution. The mean of this Poisson distribution is the expected signal $\lambda$, which is the number of background events $b$ in the data plus the expected number of gamma-ray events $\gamma$ given a source model convolved with the detector response from the detector simulations. The probability of observing a signal $N$ given an expected signal $\lambda$ is 

\begin{equation}
P(N;\lambda) = \frac{\lambda^N e^{-\lambda}}{N!}.
\end{equation}

The log likelihood of having the observation given a parameter set $\vv\theta$ in the source model is the sum of the log likelihood of each pixel in an ROI and in each \textit{f} bin, 

\begin{equation}
log\mathcal{L}(\vv{\theta} | \vv{N}) = \sum_{i}^{\mathrm{f\,bins}}\sum_{j}^{\mathrm{ROI}}(N_{ij}log\lambda_{ij}-\lambda_{ij}),
\label{equ:logL}
\end{equation}

\noindent where $N_{ij}$ and $\lambda_{ij}$ are the observed and expected signal in the $j$th pixel of the $i$th \textit{f} bin, respectively.  The term $N_{ij}!$ is discarded since it is given in the observation and independent of the parameters in the source model. The MINUIT package \cite{Minuit} is used to maximize the log likelihood by varying the parameter set $\vv{\theta}$ in the source model, i.e. to find the most likely source model given the observation.

The size of ROI is chosen to be larger than the angular resolution of the detector and possible source extent to include more photons from a given source. However, there is a high potential of source confusion for an analysis of the Galactic plane in the HAWC-111 data. In other words, it is not always possible to choose an ROI that contains only one source without contamination by photons from another source. In this case the source model may need to build with more than one source and the expected signal becomes

\begin{equation}
\lambda_{ij} = b_{ij} + \sum_{k}\gamma_{ijk},
\end{equation}

\noindent where $\gamma_{ijk}$ is the expected number of gamma ray events from the $k$th source in the $j$th pixel of $i$th \textit{f} bin and $b_{ij}$ is the background in the $j$th pixel of $i$th \textit{f} bin.

To compare different source models, a likelihood ratio test is performed. Given the log likelihood of the background-only model $log\mathcal{L}_0$ and the one-source model $log\mathcal{L}_1$.  A test statistic ($TS$) as defined in

\begin{equation}
TS = -2(log\mathcal{L}_0-log\mathcal{L}_1)
\end{equation}

\noindent indicates how much the one-source model is better over the background-only model. $TS$ follows a $\chi^2$ distribution with the degrees of freedom (DoF) being equal to the difference on the number of free parameters between two models according to Wilks' theorem. Then $TS$ can be converted to a p-value of at what level the observation is consistent with the background-only hypothesis.

\begin{figure}
\includegraphics[width=\textwidth]{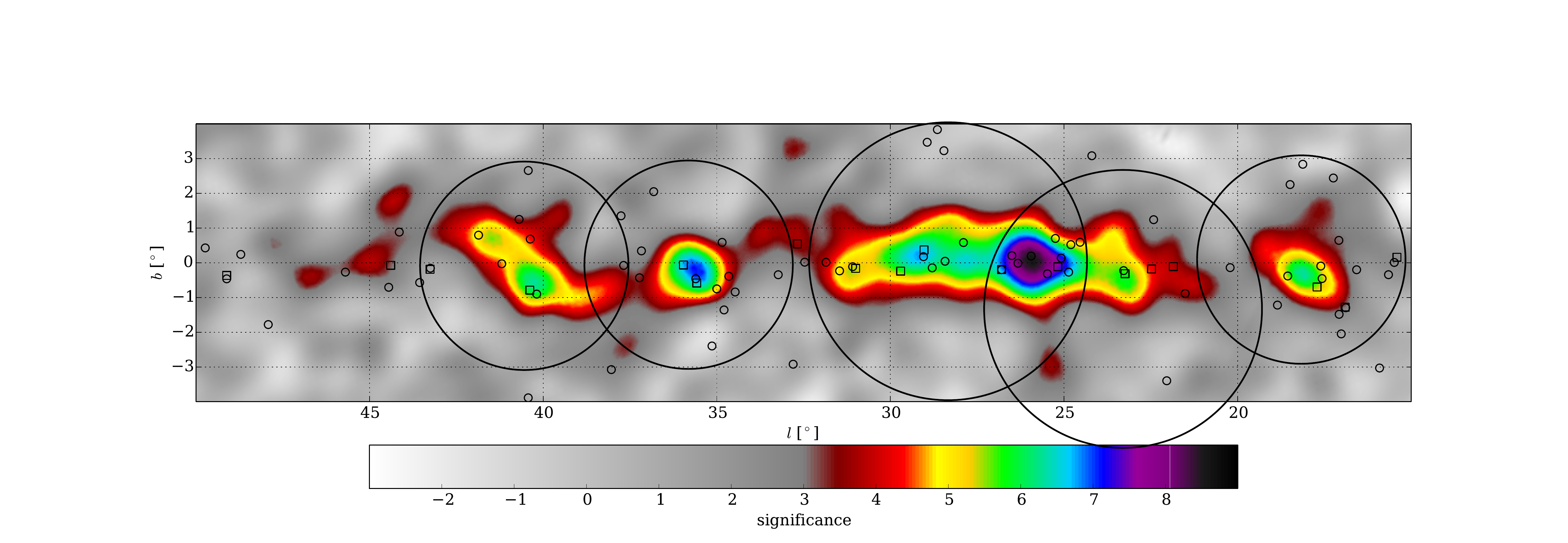}
\caption{
Significance map of the inner Galaxy region for 283 days of HAWC-111 data. 
The open squares mark TeV sources in the TeVCat \cite{tevcat} and the open circles indicate GeV sources in \textit{Fermi}-LAT 3FGL \cite{3fgl}. Five black circles show the five ROIs used in this analysis.}
\label{fig:skymap}
\end{figure}

The significance map of the inner Galaxy region as shown in Fig.\,\ref{fig:skymap} is made by moving a putative point source through each pixel, performing a maximum likelihood fit on flux normalization with spectral index fixed at 2.3, and converting the $TS$ value to significance according to Wilks' theorem. 

Due to limited sensitivity and angular resolution in this data set, all sources in the model are considered as point sources. To reduce the number of sources and number of free parameters in the likelihood fit, five ROIs are chosen in the inner Galaxy region covering all the pixels with $>5\sigma$ pre-trials significance. The position and size of each ROI, as shown in Fig.\,\ref{fig:skymap}, is carefully selected to avoid a source at the boundary. To decide the number of independent sources that are needed to properly model each ROI, an iterative likelihood ratio test is performed, in which an additional source characterized by three free parameters, right ascension, declination, and flux normalization, is added into the model at each step. This iterative procedure is repeated until a priori chosen threshold of $\Delta TS<15$ ($\sim3\sigma$ with 3 DoF) with an additional source. This procedure provides the number of sources in each ROI, and results the position and an initial estimation of the flux of each source candidate. After the iterative procedure is done for each ROI, a global fit on source fluxes in all regions is performed with fixed source positions to minimize photon contamination from sources just outside of a ROI. Finally, a single-source likelihood fit is performed on each source candidate by treating other source candidates as part of the background to obtain the flux and $TS$.

\section{Results}
Ten source candidates are identified using the likelihood analysis with $>3\sigma$ post-trials, with eight of them with tentative TeV associations. More discussion on the TeV counterparts can be found in \cite{Hui:2015}. As shown in Fig.\,\ref{fig:skymap}, a number of these source candidates are spatially coincident with GeV sources in \textit{Fermi}-LAT 3FGL or \textit{Fermi}-LAT 1FHL catalog. However, due to the limitation of angular resolution in HAWC-111 data, it is not clear if the TeV and GeV sources are associated without further investigation. The results from a joint likelihood analysis using 3ML \cite{Giacomo:2015} framework on source spectra in GeV-TeV energy range with HAWC-111 and \textit{Fermi}-LAT data will be presented at the ICRC.

\section*{Acknowledgments}
\footnotesize{
We acknowledge the support from: the US National Science Foundation (NSF);
the US Department of Energy Office of High-Energy Physics;
the Laboratory Directed Research and Development (LDRD) program of
Los Alamos National Laboratory; Consejo Nacional de Ciencia y Tecnolog\'{\i}a (CONACyT),
Mexico (grants 260378, 55155, 105666, 122331, 132197, 167281, 167733);
Red de F\'{\i}sica de Altas Energ\'{\i}as, Mexico;
DGAPA-UNAM (grants IG100414-3, IN108713,  IN121309, IN115409, IN111315);
VIEP-BUAP (grant 161-EXC-2011);
the University of Wisconsin Alumni Research Foundation;
the Institute of Geophysics, Planetary Physics, and Signatures at Los Alamos National Laboratory;
the Luc Binette Foundation UNAM Postdoctoral Fellowship program.
}

\bibliography{icrc2015-0737}

\providecommand{\href}[2]{#2}\begingroup\raggedright\begin{thebibliography}{10}

\bibitem{HESS:2006}
{\bf H.E.S.S.} Collaboration, F.~Aharonian et~al., {\it {The H.E.S.S. Survey of
  the Inner Galaxy in Very High Energy Gamma Rays}},  {\em ApJ} {\bf 636}
  (Jan., 2006) 777--797.

\bibitem{HESS:2013}
{\bf H.E.S.S.} Collaboration, S.~Carrigan et~al., {\it {The H.E.S.S. Galactic
  Plane Survey - maps, source catalog and source population}},  in {\em Proc.
  33th ICRC}, (Rio de Janeiro, Brazil), August, 2013.

\bibitem{Milagro:2007}
{\bf Milagro} Collaboration, A.~A. Abdo et~al., {\it {TeV Gamma-Ray Sources
  from a Survey of the Galactic Plane with Milagro}},  {\em ApJL} {\bf 664}
  (Aug., 2007) L91--L94.

\bibitem{Argo:2013}
{\bf ARGO-YBJ} Collaboration, B.~Bartoli et~al., {\it {TeV gamma-ray survey of
  the Northern sky using the ARGO-YBJ detector}},  {\em ArXiv e-prints} (Nov.,
  2013) [\href{http://arxiv.org/abs/1311.3376}{{\tt arXiv:1311.3376}}].

\bibitem{MAGIC:2006}
{\bf MAGIC} Collaboration, J.~Albert et~al., {\it {Observation of VHE Gamma
  Radiation from HESS J1834-087/W41 with the MAGIC Telescope}},  {\em ApJL}
  {\bf 643} (May, 2006) L53--L56.

\bibitem{VERITAS:2009}
{\bf VERITAS} Collaboration, G.~Maier et~al., {\it {VERITAS observations of
  HESS J0632+057}},  {\em ArXiv e-prints} (July, 2009)
  [\href{http://arxiv.org/abs/0907.3958}{{\tt arXiv:0907.3958}}].

\bibitem{3fgl}
{The Fermi-LAT Collaboration}, {\it {Fermi Large Area Telescope Third Source
  Catalog}},  {\em ArXiv e-prints} (Jan., 2015)
  [\href{http://arxiv.org/abs/1501.0200}{{\tt arXiv:1501.0200}}].

\bibitem{1fhl}
{\bf The Fermi-LAT} Collaboration, M.~Ackermann et~al., {\it {The First
  Fermi-LAT Catalog of Sources above 10 GeV}},  {\em ApJS} {\bf 209} (Dec.,
  2013) 34.

\bibitem{HAWC:2013}
{\bf HAWC} Collaboration, A.~U. Abeysekara et~al., {\it {Sensitivity of the
  high altitude water Cherenkov detector to sources of multi-TeV gamma rays}},
  {\em Astroparticle Physics} {\bf 50} (Dec., 2013) 26--32.

\bibitem{Pretz:2015}
{\bf HAWC} Collaboration, J.~Pretz, {\it {Highlights from the High Altitude
  Water Cherenkov Observatory}},  in {\em Proc. 34th ICRC}, (The Hague, The
  Netherlands), August, 2015.

\bibitem{Patrick:2015}
{\bf HAWC} Collaboration, P.~Younk et~al., {\it {A High-level Analysis
  Framework for HAWC}},  in {\em Proc. 34th ICRC}, (The Hague, The
  Netherlands), August, 2015.

\bibitem{Paco:2015}
{\bf HAWC} Collaboration, F.~Salesa, {\it {Observations of the Crab Nebula with
  Early HAWC Data}},  in {\em Proc. 34th ICRC}, (The Hague, The Netherlands),
  August, 2015.

\bibitem{Calibration:2015}
{\bf HAWC} Collaboration, F.~Salesa et~al., {\it {The Calibration System of the
  HAWC Gamma-Ray Observatory}},  in {\em Proc. 34th ICRC}, (The Hague, The
  Netherlands), August, 2015.

\bibitem{Healpix:2005}
K.~M. {G{\'o}rski}, E.~{Hivon}, A.~J. {Banday}, B.~D. {Wandelt}, F.~K.
  {Hansen}, M.~{Reinecke}, and M.~{Bartelmann}, {\it {HEALPix: A Framework for
  High-Resolution Discretization and Fast Analysis of Data Distributed on the
  Sphere}},  {\em ApJ} {\bf 622} (Apr., 2005) 759--771.

\bibitem{Milagro:2003}
R.~Atkins et~al., {\it {Observation of TeV Gamma Rays from the Crab Nebula with
  Milagro Using a New Background Rejection Technique}},  {\em ApJ} {\bf 595}
  (Oct., 2003) 803--811.

\bibitem{Corsika:1998}
D.~Heck et~al., {\it {CORSIKA: A Monte Carlo Code to Simulate Extensive Air
  Showers}},  Tech. Rep. FZKA-6019, 1998.

\bibitem{Geant4:2003}
{\bf GEANT4} Collaboration, S.~Agostinelli et~al., {\it {GEANT4: A simulation
  toolkit}},  {\em Nucl. Instrum. Meth.} {\bf A506} (2003) 250--303.

\bibitem{Minuit}
F.~{James} and M.~{Roos}, {\it {Minuit - a system for function minimization and
  analysis of the parameter errors and correlations}},  {\em Computer Physics
  Communications} {\bf 10} (Dec., 1975) 343--367.

\bibitem{tevcat}
``Online catalog for tev astronomy.'' http://tevcat.uchicago.edu.

\bibitem{Hui:2015}
{\bf HAWC} Collaboration, C.~M. Hui and H.~Zhou, {\it {HAWC Observations of
  Supernova Remnants and Pulsar Wind Nebulae}},  in {\em Proc. 34th ICRC}, (The
  Hague, The Netherlands), August, 2015.

\bibitem{Giacomo:2015}
G.~Vianello et~al., {\it {The Multi-Mission Maximum Likelihood framework
  (3ML)}},  in {\em Proc. 34th ICRC}, (The Hague, The Netherlands), August,
  2015.

\end{thebibliography}\endgroup

\end{document}